\documentclass[12pt,a4paper]{article}

\usepackage{ulem}
\usepackage{amsmath,amssymb,amsthm,amsfonts,makeidx}
\usepackage{color}
\usepackage{graphicx}
\usepackage{enumerate}
\usepackage{mathrsfs}
\usepackage{setspace}
\IfFileExists{url.sty}{\usepackage{url}}
                     {\newcommand{\url}{\text}}
                     \usepackage{anysize}

\newtheorem{lem}{Lemma}[section]
\newtheorem{theo}{Theorem}[section]
\newtheorem{remark}{Remark}[section]

\begin{document}


\begin{center}
{\Large \textbf{Interpretable Clustering using Unsupervised Binary Trees}}

\

{\large Ricardo Fraiman $^{\ast}$, Badih Ghattas $^{\ast\ast}$ and Marcela Svarc $^{\ast\ast\ast}${\footnote{Corresponding author:
Marcela Svarc, Departamento de Matem\'aticas y Ciencias, Universidad  de San Andr\'es, Vito Dumas 284, Victoria (1644), Buenos Aires, Argentina. Email: msvarc@udesa.edu.ar}}}

\

\noindent $^{\ast}$Universidad de San Andr\'es, Argentina and Universidad de la Rep\'ublica, Uruguay.
\\[0pt]$^{\ast\ast}$ Universit\'e de la M\'editerrann\'ee Faculté des Sciences de Luminy, Département de Mathématiques
          163 Avenue de Luminy, 13288 Marseille cedex 09, France.
\\[0pt]$^{\ast\ast\ast}$ Universidad de San Andr\'es and CONICET, Argentina.

\end{center}

%


\begin{abstract}
We herein introduce a new method of interpretable clustering that uses unsupervised binary
trees. It is a three-stage procedure, the first stage of which entails a series of recursive binary splits to reduce  the heterogeneity of
the data within the new subsamples. During the second stage (pruning), consideration is given to whether adjacent nodes can be aggregated. Finally, during the third stage (joining), similar clusters are joined together, even if they do not share the same parent originally. Consistency results are obtained, and the procedure is used on simulated and real data sets.

\end{abstract}

\noindent {\em Running Title:} Clustering a la CART.

\section{Introduction}
Clustering is a means of unsupervised classification, is a common
technique for the  statistical analysis of data, and has applications in many
fields, including medicine, marketing and economics, among other areas.
The term ``cluster analysis" (first used by Tryon, \cite{Tryon})
includes the use of any of a number of different algorithms and methods for grouping
similar data into a number of different categories. The grouping is achieved
in such a way that ``the degree of association between data is at a maximum
if the data belong to the same group and at a minimum otherwise".

Cluster analysis or clustering involves the assignment of a set of observations from $\mathbb{R}{^p}$ into subsets (called clusters), such that observations in the same cluster are similar in ``some sense". The definitions are quite vague, in that there is no clear objective function of the population that can be used to measure the performance of a clustering procedure. Implicit in each clustering algorithm is an objective function that varies from one method to another. It is important to note that although most clustering procedures require the number of clusters to be known beforehand, generally in practice it is not.

In contrast, in supervised classification the number of groups is known and we additionally have both a learning sample and a universal objective function, i.e. to minimise the number of misclassifications, or in terms of the population, to minimise the Bayes error.

Despite these differences, there are a number of similarities between supervised and unsupervised classification. Specifically, there are many algorithms that share the same spirit in both cases.

Algorithms that use supervised and unsupervised classification \cite{Hastie} can either be partitional or hierarchical. Partitional algorithms determine all the groups at once. The most widely used and well-studied partitioning procedure for cluster analysis is that of the k-means algorithm. Hierarchical algorithms successively identify groups that split from or join groups that were established previously. These algorithms can either be agglomerative (``bottom-up") or divisive (``top-down"). Agglomerative algorithms begin with each element as a separate group and merge them into successively larger groups. In contrast, divisive algorithms begin with the whole set and proceed to split it into successively smaller groups. Hierarchical algorithms create a hierarchy of partitions that may be represented in a tree structure. The best known hierarchical algorithm for supervised classification is CART \cite{Breiman}.

CART has a further property of interest. The partition tree is built using a few binary conditions obtained from the original coordinates of the data. In most cases, the interpretation of the results may be summarised in a tree that has a very simple structure. The usefulness of such a scheme of classification is valuable not only for the rapid classification of new observations, but it can also often yield a much simpler ``model" for explaining why the observations are classified in a particular group, a property that is remarkably important in many applications. Moreover, it is important to stress that the algorithm assumes no kind of parametric model for the underlying distribution.

Recently, several new methods have been proposed for clustering analysis, see for instance Garcia Escudero et al. \cite{graciaescudero} for a review with focus on  robust clustering procedure. Other recent proposals have been made by Pe\~na and Prieto  (\cite{penaprieto01},\cite{penaprieto06}), Fraley and Raftery \cite{fraleyraftery02}, Oh and Raftery \cite{ohraftery}, Walther \cite{walther}, among others. But just a few different methods using decision trees to obtain clusters have previously been proposed. Liu et al. \cite{LiuXiaYu} use decision trees to partition the data space into clusters and empty (sparse) regions at different levels of detail. Their method uses the idea of adding an artificial sample of size $N$ that is uniformly distributed over the space. With these $N$ points added to the original data set, the problem then becomes one of obtaining a partition of the space into dense and sparse regions. Liu et al. \cite{LiuXiaYu} treat this problem as a classification problem that uses a new ``purity" function that is adapted to the problem and is based on the relative densities of the regions concerned.

Chavent et al. \cite{chavent} obtained a binary clustering tree that applies to a particular variable and its binary transformation. They presented two alternative procedures. In the first, the splitting variables are recursively selected using correspondence analysis, and the resulting factorial scores lead to the binary transformation. In the second, the candidate variables and their variable transformations are simultaneously selected by a criterion of optimisation in which the resulting partitions are evaluated. Basak et al. \cite{Basak} proposed four different measures for selecting the most appropriate characteristics for splitting the data at every node, and two algorithms for partitioning the data at every decision node. For the specific case of categorical data, Andreopoulus et al. \cite{Andre} introduced HIERDENC, which is an algorithm that searches the dense subspaces on the ``cube" distribution of the values presented in the data at hand.

Our aim herein is to propose a simple clustering procedure that has the same appealing properties as CART. We introduce the hierarchical top-down method of CUBT (Clustering using unsupervised binary trees), in which the clustering tree is based on binary rules on the original variables, and this will help us to understand the structure of the clustering.

 There are three stages in our procedure. In the first, we grow a maximal tree by applying a recursive partitioning algorithm. In the second, we prune the tree using a criterion of minimal dissimilarity. In the final stage, we aggregate the leaves of the tree that do not necessarily share the same direct ascendant.

We do not claim that the new method we introduce is always more efficient nor better than others.  For each particular problem, depending on the cluster structure some methods behave better than others, and quite often the difference in efficiency is just a small (but important) improvement. This is also the case in supervised classification. CART is far from being the best universally more efficient algorithm. However it has a quite good behavior for a large class of classification problems.
We will show along the paper, that the main advantages of our clustering method (mostly shared with CART for classification problems) are the following:
\begin{itemize}
\item [a)] \bf  Flexibility. \it The method is able to perform good clustering for a large family of cluster structure. \rm As long as the true clusters can be separated by some partition built as the intersection of a arbitrarily large finite number of half--spaces, whose boundaries are orthogonal to the original coordinates system the method will work properly.
\item [b)] \bf Interpretability. \it The final partition is explained in terms of binary partitions on the original variables. \rm This property is fundamental for many applications.
    \item [c)] \bf Efficiency. \it We get a good performance in terms of correct clustering allocation for a large family of clusters structures. \rm
    \item [d)] \bf Population version. \it We provide a population version of the final partition, regarded as a random vector $\mathbf{X} \in \mathbb{R}^p$ with unknown distribution $P$. We then show that the algorithm (the empirical version) converges a.s. to the population final partition.  \rm This kind of property is essential in statistics in order to understand well when and why a method will be adequate, by looking at the population version. This is briefly discussed on Section 2.3.
         \end{itemize}


The remainder of the paper is organised as follows. In Section 2, we introduce some notation and we describe the empirical and population versions of our method. The latter describes the method in terms of the population, regarded as a random vector $\mathbf{X} \in  \mathbb{R}^p$ with an unknown distribution $P$. The consistency of our method is described in Section 3. In Section 4, we present the results of a series of simulations, in which we used our method to consider several different models and compare the results with those produced by the k-means algorithm, MCLUST and DBSCAN. Using a synthetic data set, we also compared the tree structures produced by CART (using the training sample, with the labels) and CUBT, considering the same sample in each case without the labels. A set of real data is analysed in Section 5, and our concluding remarks are made in Section 6. All proofs are given in the Appendix.

\section{Clustering a la CART}

We begin by establishing some notation. Let $\mathbf{X} \in  \mathbb{R}^p$ be a random p-dimensional real vector with coordinates $X(j), j=1,\dots,p$, such that $\mathbb{E}(\Vert
\mathbf{X} \Vert ^2) < \infty$. The data consist in $n$ random independent and identically distributed realisations of $\mathbf{X}$, $\Xi=\{\mathbf{X}_1, \ldots, \mathbf{X}_n\}$. For the population version the space is $\mathbb{R}^p$, while for the empirical version the space is $ \Xi$. We denote the nodes of the tree by $t$. Each node $t$ determines a subset of $\mathbb{R}^p$ , $t \subset \mathbb{R}^p$. We assign the whole space to the root node.

Even though our procedure uses the same general approach as CART in many respects, two main differences should be stressed. First, because we use unsupervised classification, only information about the observations without labels is available. Thus the splitting criterion cannot make use of the labels, as it can in CART.
The second essential difference is that instead of having one final pruning stage, in our algorithm we subdivide this stage into two, in that we first prune the tree and then use a final joining process. In the first of these two procedures, we evaluate the merging of adjacent nodes, and in the second the aim is to aggregate similar clusters that do not share the same direct ascendant in the tree.

\subsection{Forward step: maximal tree construction} \label{split}

Because we are using a top-down procedure, we begin by assigning the whole space to the root node. Let $t$ be a node and $\hat{t}= \Xi \cap t$, the set of observations obtained from the sample concerned. At each stage, a terminal node is considered to be split into two subnodes, namely the left and right child, $t_l$, $t_r$, if it fulfills a certain condition. At the beginning there is only one node, i.e. the root, which contains the whole space. The splitting rule has the form $x(j) < a$, where $x(j)$ is a variable and $a$ is a threshold level. Thus,
$t_l = \{x \in \mathbb{R}^p: x(j) \leq a\}$ and $t_r = \{x \in \mathbb{R}^p: x(j) > a\}$.

Let $X_t$ be the restriction of $X$ to the node $t$, i.e. $X_t = \mathbf{X} |\{\mathbf{X} \in t\}$, and $\alpha_t$ the probability of being in $t$, $\alpha_t = P(\mathbf{X} \in t)$. $R(t)$ is then a heterogeneity measure of $t$ and is defined by,
\begin{equation}
R(t) = \alpha_t \; trace(Cov(X_t)),\label{deviance}
\end{equation}
where, $cov(X_t)$ is the covariance matrix of $X_t$. Thus, $R(t)$ is an approximate measure of the ``mass concentration" of the random vector $X$ at set $t$, weighted by the mass of set $t$. In the empirical algorithm $\alpha_t$ and $Cov(X_t)$ are replaced by their empirical versions (estimates), and $R(t)$ is called the
{\it deviance}. We then denote  $n_t$ to be the cardinal of the set $\hat{t}$, $n_{t}=\sum_{i=1}^{n}\mathcal{I}\left\{ \mathbf{X}_{i}\in
t\right\}$, (where $\mathcal{I}_A$ stands for the indicator function of  set $A$), and hence the estimated probability is $\widehat{\alpha }_{t}=\frac{n_{t}}{n}$, the estimate of $E\left( \left\Vert X_{t}-\mu _{t}\right\Vert ^{2}\right)$ is
\begin{equation*}
\frac{\sum_{\left\{ \mathbf{X}_{i}\in t\right\} }\left\Vert \mathbf{X}_{i}-\overline{X}%
_{t}\right\Vert ^{2}}{n_{t}},
\end{equation*}
where $\overline{X}_{t}$ is the empirical mean of the observations on $t$ and the estimate of the deviance is,
\begin{equation}
 \hat{R}(t)=\frac{\sum_{\left\{ \mathbf{X}_{i}\in t\right\} }\left\Vert \mathbf{X}_{i}-\overline{X}
_{t}\right\Vert ^{2}}{n}.
\end{equation}

The best split for $t$ is defined as the couple
$(j,a)\in \{1,\dots,p\} \times \mathbb{R}$, (the first element indicates the variable where the partition is defined and the second is the threshold level) that maximises,
\begin{equation}
 \Delta(t,j,a) = R(t)- R(t_l) - R(t_r).
\label{Delta}
\end{equation}
It is easy to verify that $\Delta(t,j,a) \ge 0$ for all $t,j,a$, and this property is also verified by all the splitting criteria proposed in CART.

\begin{remark} (\textbf{Uniqueness}) As in CART the maximum in (\ref{Delta}) may not by unique. Typically, the maximum is attained only for one variable, $j$, but there may be many values $a$ for which it is attained, for instance a union of intervals. In those cases we will choose the smallest value $a$ for which the maximum is attained. More precisely, for fixed $j$, $a$ is defined as $$\inf {\{argmax_{a \in \mathbb{R}} \Delta(t,j,a)\}}.$$ If there are several variables that attain the maximum then we choose the variable $j$ with smallest index, and then if needed we choose the smallest value $a$ where the maximum is attained.
\end{remark}

 We fix two parameters  $ \tau, mindev \in (0,1)$.

We begin with the whole space being assigned to the root node, and each node is then split recursively until one of the following stopping rules is satisfied:  (i) $\alpha_t < \tau$; (ii) The reduction in deviance is less than $mindev \times R(\mathcal{S})$, where $\mathcal{S}$ is the whole space. For the empirical version of the algorithm we replace $\alpha_t$ by $\hat{\alpha}_t$ and $\Delta(t,j,a)$ by $\hat{\Delta}(t,j,a) = \hat{R}(t) - \hat{R}(t_l) - \hat{R}(t_r)$, and denote by $minsize= [\tau n]$ (the minimum size of a node).
$minsize$ and $mindev$ are tuning parameters that must be supplied by the user. The uniqueness problem also appears on the empirical version, and it can be treated as in the population algorithm.

When the algorithm stops, a label is assigned to each leaf (terminal node), and we then call the actual tree the maximal tree.
At this point, we have obtained a partition of the space and in consequence a partition of the data set, in which each leaf is associated with a cluster. Ideally, this tree has at least the same number of clusters as the population, although in practice it may have too many clusters, and then an agglomerative stage must be applied as in CART.
It is important to note if the number of clusters $k$ is known, the number of leaves should be greater or equal to $k$.   Small values of $mindev$ ensure a tree that has many leaves. Moreover, if the tree has the same number of leaves as the number of clusters, it is not then necessary to run the subsequent stages of the algorithm.




\subsection{Backward step: pruning and joining}

In the next step, we successively use two algorithms to give the final conformation of the grouping. The first prunes the tree and the second merges non-adjacent leaves (we herein refer to this as ``joining").
We now introduce a pruning criterion that we refer to as ``minimum dissimilarity pruning".

\subsubsection{Minimum dissimilarity pruning}

In this stage, we define a measure of the dissimilarity between sibling nodes, and then collapse these nodes if this measure is lower than a certain threshold. We first consider the maximal tree $T_0$ obtained in the previous stage. Let $t_l$ and $t_r$ be a pair of terminal nodes that share the same direct ascendant. We then define (in population terms) the random variables $W_{l(r)}=D(X_{t_l}, sop(X_{t_r}))$, (where $sop(Z)$ stands for the support of the random variable $Z$) as the Euclidian distance between the random elements of $X_{t_l}$ and the support of $X_{t_r}$ and respectively $W_{r(l)}=D(sop(X_{t_l}), X_{t_r})$. For each of them we define a dissimilarity measure
\begin{eqnarray}
\Delta_{l(r)} = \int _0^{\delta} q_{\nu}(W_{l(r)})d\nu \nonumber, \\
\Delta_{r(l)} = \int _0^{\delta} q_{\nu}(W_{r(l)})d\nu \nonumber, \\
\end{eqnarray}
where $q_{\nu}(W_{lr})$ stands for the quantile function, ($\mathbb P(W_{l(r)} \leq q_{\nu})=\nu$) and  $\delta$ is a proportion,  $\delta
\in (0,1)$. \\
Finally we consider as a measure of dissimilarity between the sets $t_l$ and $t_r$
$$
\Delta_{lr} = max\{{\Delta_{l(r)},\Delta_{r(l)}}\}.
$$
If $\Delta_{lr} < \epsilon$, we prune at the node $t$, i.e. we replace $t_l$ and $t_r$ by $t_l \cup t_r$ in the partition.

 Observe that since $\Delta_{l(r)}$ and $\Delta_{r(l)}$ are averages of small quantiles of $D(X_{t_l}, sop(X_{t_r}))$ and of $D(X_{t_r}, sop(X_{t_l}))$ respectively, $\Delta_{lr}$ can be thought as ``a more resistant version" of the distance between the supports of the random vectors  $X_{t_l}$ and $X_{t_r}$.\\

We consider the natural plug-in estimate for the dissimilarity measure $\Delta$, that is defined as follows.
 Let $n_l$ (resp. $n_r$) be the size of $\hat t_l$ (resp. $\hat t_r$).
We may then consider for every $x_i\in \hat t_l$ and $y_j \in \hat t_r$,  the sequences,
$ \tilde{d}_i =  min_{y\in \hat t_l}d(x_i,y), \;\; \tilde{d}_j =  min_{x\in \hat t_r}d(x,y_j)$
and their ordered versions, denoted as $d_i$ and $d_j$. For $\delta \in [0,1]$, let
$$ \bar{d}_l^{\delta} = \frac{1}{\delta n_l}\sum_{i=1}^{\delta n_l} d_i,\;\;\;  \bar{d}_r^{\delta} = \frac{1}{\delta n_r}\sum_{j=1}^{\delta n_r} d_j.$$

We compute the dissimilarity between $t_l$ and $t_r$ to be,
$$d^{\delta}(l,r)= d^{\delta}(t_l,t_r)  = max(\bar{d}_l^{\delta},\bar{d}_r^{\delta}),$$
and at each stage of the algorithm the leaves, $t_l$ and $t_r$, are merged into the ascendant node $t$ if
$d^{\delta}(l,r) \leq \epsilon$
where $\epsilon >0$. The dissimilarity pruning makes use of the two parameters $\delta$ and  $\epsilon$, which we hereafter refer to as ``mindist".

\subsubsection{Joining} \label{joining}

\begin{figure}
\begin{center} \includegraphics[height=4cm,width=4cm]{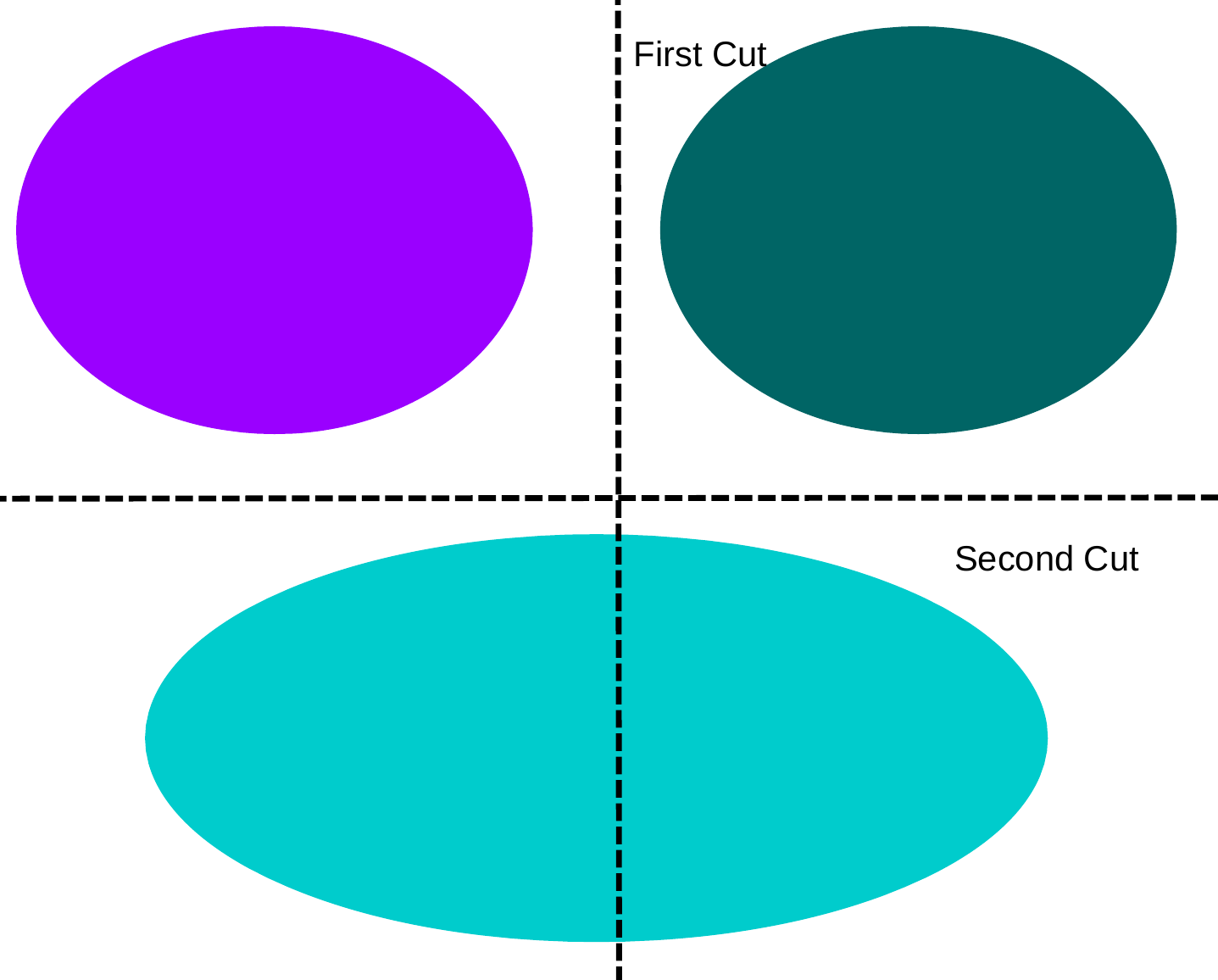}  \end{center}
 \caption{The lower group is split on the first stage and it cannot be merged in the pruning stage, but will be when using the joining step.}
 \label{divisiones}
\end{figure}

The aim of the joining step is to aggregate those nodes that do not share the same direct ascendent. The criterion used in the joining step is
the same as the used on the pruning one without the restriction of being sibling nodes. The need of this step is illustrated in Figure \ref{divisiones}.

Here, all pairs of terminal nodes $t_i$ and $t_j$ are compared by computing,
$d^{\delta}(i,j)$. As in standard hierarchical clustering, pairs of terminal nodes are successively aggregated starting from the pair $i,j$ with minimum   $d^{\delta}(i,j)$ value, thereby producing one less cluster at each step.
We consider two different stopping rules for the joining procedure, which correspond to the two cases where the number of clusters $k$ is either known or unknown. We denote by $m$ the number of leaves of the pruned tree.
If $k$ is known, the following step is repeated until $m \leq k$:
\begin{itemize}
\item For each pair of values  $(i,j)$ , $1 \leq i < j \leq m$, let $(\tilde{i},\tilde{j})= arg\min_{i,j} \{d^{\delta}(i,j)\}$. Replace $t_{\tilde{i}}$ and $t_{\tilde{j}}$ by its union $t_{\tilde{i}} \cup t_{\tilde{j}}$, put $m=m-1$ and proceed.
\end{itemize}
If $k$ is unknown,
\begin{itemize}
\item if $d^{\delta}_{\tilde{i} \tilde{j}} < \eta$ replace $t_{\tilde{i}}$ and $t_{\tilde{j}}$ by its union $t_{\tilde{i}} \cup t_{\tilde{j}}$, where $\eta>0$ is a given constant, and continue until  this condition does not apply.
\end{itemize}
In the first case, the stopping criterion is the number of clusters, while in the second case a threshold value of $\eta$ for $d^{\delta}(i,j)$ must
be specified.\\

\subsection{CUBT and $k$-means}

In the following section we discuss informally the circumstances in which both our procedure and the well-known k-means algorithm should produce a reasonably good set of results.
We shall consider those cases where there are ``nice groups" that are strictly separated. More precisely,  let $A_1, \ldots, A_k$ be disjoint connected compact sets on $ \mathbb{R}^p$ such that $A_i = \overline{A_i^0}$ for $i=1, \ldots, k$, and $\{P_i: i=1, \ldots, k\}$ their probability measures on $ \mathbb{R}^p$ with supports $\{A_i:  i=1, \ldots, k\}$ .

A typical case may be obtained by defining a random vector $\mathbf{X}^*$ with a density $f$ and then considering the random vector $\mathbf{X} = \mathbf{X}^*\vert \{f > \zeta\}$  for a positive level set $\zeta$, as in a number of hierarchical clustering procedures.

An admissible family for CUBT is the family of sets  $A_1, \ldots, A_k$ such that there exist  another family of disjoint sets $B_1, \ldots, B_k$ that are built up as the intersection of a finite number of half-spaces delimited by hyperplanes that are orthogonal to the coordinate axis that satisfying  $A_i \subset B_i$.

In contrast, the k-means algorithm is defined through the vector of centres $(c_1 \ldots, c_k)$ that minimise
$\mathbb{E}\left(\min_{j=1, \ldots,k} \Vert \mathbf{X} - c_j \Vert \right)$.
Associated with each centre $c_j$ is the convex polyhedron $S_j$ of all points in $\mathbb{R}^p$ that are closer to $c_j$ than to any other center, called the Voronoi cell of $c_j$.  The sets in the partition $S_1, \ldots, S_k$ are the population clusters for the $k$--means algorithm. The population clusters for the $k$-means algorithm are therefore defined by exactly $k$ hyperplanes in an arbitrary position.

Then, an admissible family for the $k$-means algorithm will be a family of sets $A_1, \ldots, A_k$ that can be separated by exactly $k$ hyperplanes.

Although the hyperplanes can be in an arbitrary position, no more than $k$ of them can be used.

It is clear that in this sense CUBT is much more flexible than the k-means algorithm, because the family of admissible sets is more general. For example, $k$-means will necessarily fail to identify nested groups, while CUBT will not.

Another important difference between $k$--means and CUBT is that our proposal is less sensitive to small changes in the parameters that define the partition. Effectively, small changes in these will produce small changes in the partition. However, small changes in the centres $(c_1 \ldots, c_k)$ that define the $k$-means partition can produce significant changes in the associated partition as given by the Voronoi cells.

\begin{figure}[!ht]
\begin{center} \includegraphics[width=10cm]{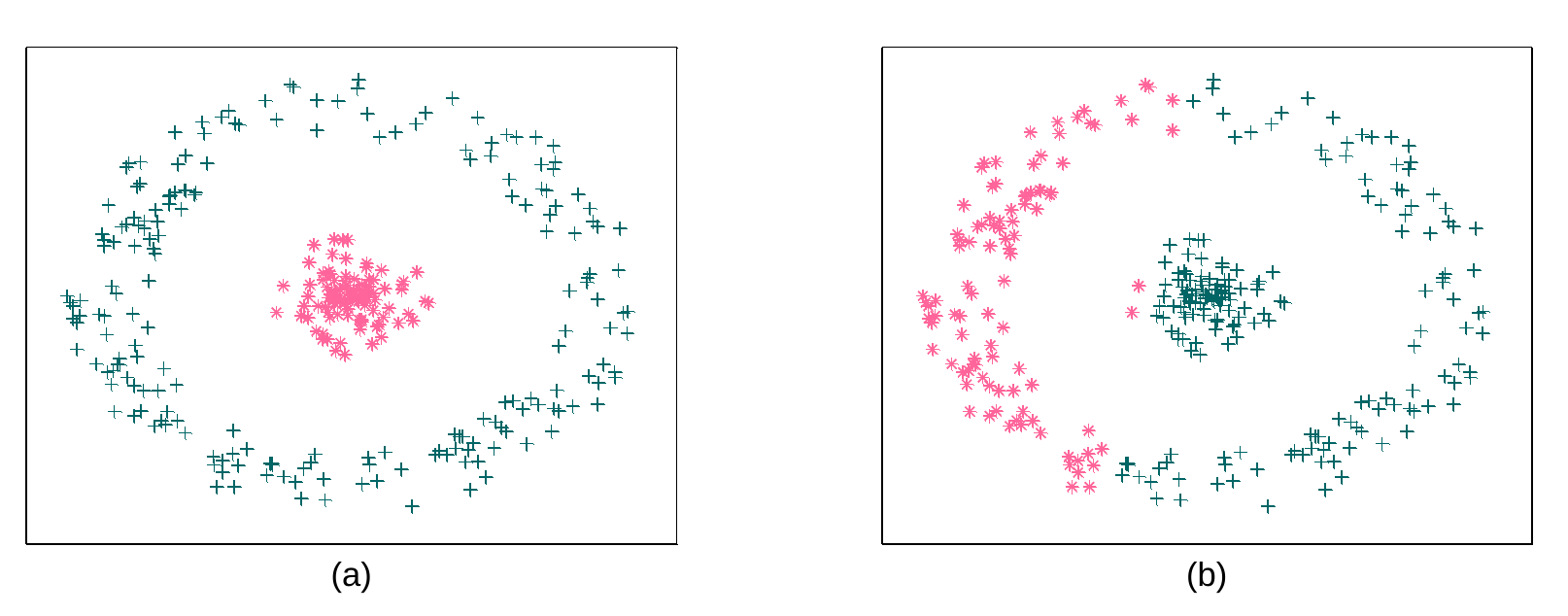}   \end{center}
 \caption{(a) CUBT cluster allocation. (b) k-means cluster allocation.}
 \label{ejanindados}
\end{figure}

Figure \ref{ejanindados} show an example where CUBT has a good performance and clearly $k$--means does not work.

\section{Consistency of CUBT}

In the following section we present some theoretical results on the consistency of our algorithm. We first prove an important property, that of the monotonicity of the deviance with increasing tree size.\\
A simple equivalent characterisation of the function $R(t)$ is given in the following Lemma.
\begin{lem}
 Let $t_l$ and $t_r$ be disjoint compact sets on $ \mathbb{R}^p$ and $ \mu_s =  \mathbb{E}(X_{t_s}),  \  \  s=l,r$. If $t = t_l \cup t_r$ we have,
 \begin{equation}
  R(t) = R(t_l)+R(t_r) + \frac{\alpha_{t_l} \alpha_{t_r}}{\alpha_t}  \Vert \mu_l - \mu_r\Vert^2.
  \label{deltaR}
 \end{equation}
 \label{lem1}
\end{lem}
The proof is given in the Appendix.

\begin{remark} { \bf Monotonicity of the function $R(.) $ and geometric interpretation.} Observe that Lemma (\ref{lem1}) entails that for all disjoint compact sets $t_l, t_r$ and $t=t_l\cup t_r$, the function $ R(.)$ is monotonic in the sense that, $R(t) \geq R(t_l) + R(t_r)$.
Moreover, $R(t)$ will be close to $R(t_l)+R(t_r)$ when the last term on the right hand side of equation (\ref{deltaR}) is small. This will occur either if one of the sets $t_l$ and $t_r$ has a very small fraction of the mass of $t$ and/or if the centers of the two subsets $t_l,t_r$ are very close together. In either cases we do not want to split the set $t$.
\end{remark}

The following results show the consistency of the empirical algorithm in relation to its population version. We begin with the splitting algorithm and then follow this by pruning and joining.
\begin{theo}
Assume that the random vector $\mathbf{X}$ has distribution $P$ and a density $f$ that fulfils the condition that
 $\Vert x \Vert ^2 f(x)$ is bounded. Let $\mathbf{X}_1, \ldots, \mathbf{X}_n$ be iid random vectors with the same distribution as $\mathbf{X}$ and denote by $P_n$ the empirical distribution of the sample $\mathbf{X}_1, \ldots, \mathbf{X}_n$.\\
 Let $ \{t_{1n}, \ldots, t_{m_{n}n} \}$ be the empirical binary partition obtained from the forward empirical algorithm, and $ \{t_{1}, \ldots, t_m \}$ the population version. We then find that ultimately, i.e. there exists $n_0$ such that for $n > n_0$, ,  $m_n = m$  and  each pair $(i_{jn}, a_{jn})  \in   \{1,  \ldots, p \} \times  \mathbb{R}$ in the determination of the empirical partition converges a.s. to the corresponding one  $(i_{j}, a_{j})  \in   \{1,  \ldots, p \} \times  \mathbb{R}$ for the population values. In particular, it implies that, $\lim_{n  \to  \infty} \sum_{i=1}^m  P \left( t_{in}  \Delta t_i \right) = 0$, where $\Delta$ stands for the symmetric difference.

\label{theo1}
\end{theo}

The proof is given in the Appendix.

\begin{theo}
 Let $ \{t^*_{1n}, \ldots, t^*_{k_{n}n} \}$ be the final empirical binary partition obtained after the forward and backward empirical algorithm has been applied, and $ \{t^*_{1}, \ldots, t^*_k \}$ be the population version. Under the assumptions of Theorem \ref{theo1} we ultimately have that $k_n=k $  ($k_n=k$   for all $n$ if $k$ is known), and $\lim_{n  \to  \infty} \sum_{i=1}^k  P \left( t^*_{in}  \Delta t^*_i \right) = 0$.
\label{theo2}
\end{theo}
The proof is given in the Appendix.

\section{Some experiments. \label{simulations}}

In the following Section, we present the results of a simulation in which we tested our method using four different models. We consider separately the cases where the number of groups is known and where it is unknown. If the number of groups is known we compare the results obtained with those found using the $k$-means algorithm. It is well known the performance of the $k$-means algorithm depends strongly on the position of the centroids initially used to start the algorithm, and a number of different methods have been proposed to take account of this effect (see Steinley, \cite{Steinley}). We herein follow the recommendations made in this last reference, and consider ten random initialisations, keeping the one with the minimum within-cluster sum of squares given by, $\sum_{i=1}^{n} \sum_{j=1}^{k}\Vert \mathbf{X}_{i}-c_{j}\Vert ^{2}\mathcal{I}_{\{ \mathbf{X}_{i}\in G_{j}\}}$, where $G_j$ is the $j$--th group and $c_j$ is the corresponding center. We refer to this version as  $k$-means$(10)$.

 We also compare our results with the  well known  model--based clustering procedure proposed by Fraley and Raftery (\cite{fraleyraftery02}, \cite{fraleyraftery09}). The MCLUST algorithm assumes that the sample has been generated from a mixture of $G$ normal distributions, with ellipsoidal
 covariance matrices with variable shape, volume and orientation, and estimates the parameters of each population considering an EM algorithm and the most appropriate model is selected by means of the Bayesian Information Criteria. This procedure is versatile and it can be applied whether the number of groups is stated or not, hence
 we report the results for both cases.

We consider the algorithm DBSCAN (Density Based Spatial Clustering of Applications with Noise, \cite{ester}) which is suitable for discovering arbitrarily shaped clusters, since clusters are defined as dense regions separated by low--density regions. This algorithm estimates the number of clusters, as most of the density based algorithms do, hence we compare the results with those where the number of groups is unknown. The algorithm depends on two parameters, the number of objects in a neighborhood of an object, our input was 5, and the neighborhood radius, the authors, Ester et al \cite{ester}, recommend to avoid putting an input if this parameter is unknown.


\subsection{Simulated data sets}

We herein consider four different models (M1 - M4) having different number of groups and dimensions. For each model clusters have equal number of observations.

\begin{figure}[!ht]
  \includegraphics[height=3.5cm,width=10cm]{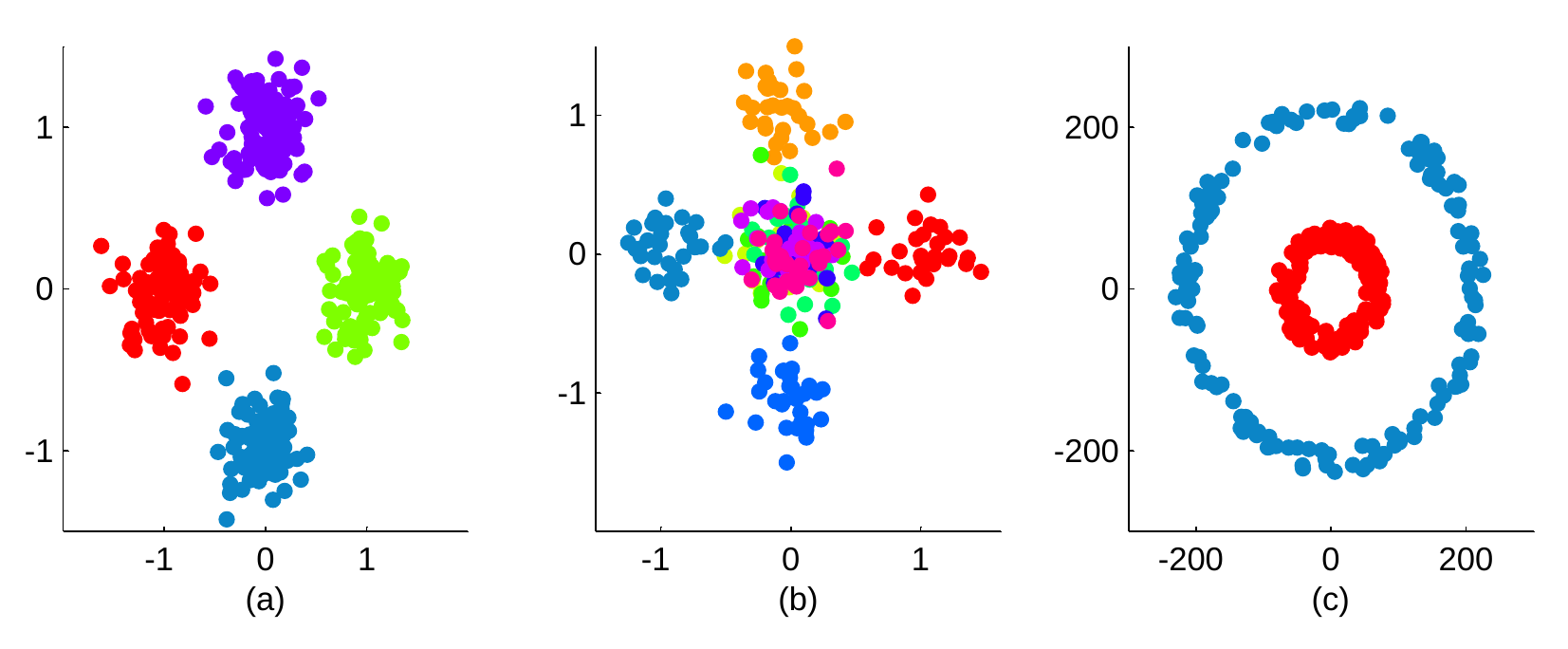}
 \caption{(a) Scatter plot corresponding to M1 for $\sigma=0.19$ (b) Two-dimensional projection scatter plot corresponding to M2  for $\sigma=0.19$ (c) Scatter plots corresponding to M3}
 \label{M123}
\end{figure}

\begin{itemize}

%
%

\item [M1.]  Four groups in dimension 2 with 100 observations per group. The data are generated using : $N(\mu_i, \Sigma)$, $i=1, \ldots 4$, and distributed with centers
$(-1,0), (1,0), (0,-1), (0,1)$ and covariance matrix $ \Sigma = \sigma^2 Id$ with $\sigma=0.11, 0.13, 0.15,0.17,0.19.$
Figure \ref{M123} (a) gives an example of the data generated using model M1 with $\sigma=0.19.$

\item [M2.] Ten groups in dimension 5 with 30 observations per group. The data are generated using $N(\mu_i, \Sigma), i=1, \ldots, 10$. The means $\mu_i$ of the first five groups, are the vectors of the canonical basis $e_1, \ldots, e_5$ respectively, while the centers of the five remaining groups  are
$ \mu_{5+i} =- e_i,  \  \ i=1,  \ldots 5$. In all cases, the covariance matrix is  $\Sigma = \sigma^2 Id$, $\sigma=0.11,0.13,0.15,0.17$.
Figure \ref{M123} (b) gives an example of data generated using model M4 with $\sigma=0.19$ projected on the two first dimensions.


\item [M3.] Two groups in dimension 2 with 150 observations per group. The data are uniformly generated on two concentric rings. The radius of the inner ring is between 50 and 80 and the radius of the outer ring is between 200 and 230.
    Figure \ref{M123} (c) gives an example of data generated using model M3.

\item [M4.] Three groups in dimension 50 with 25 observations per group. The data are generated using $N(\mu_i, \Sigma), i=1, 2, 3$. The mean of the first group is $(0.1,\dots,0.1),$ the second group is centered at the origin and the last group has mean $(-0.1,\dots,-0.1)$. In all cases, the covariance matrix is  $\Sigma = \sigma^2 Id$, $\sigma=0.03$ or $0.05$.

\end{itemize}

\subsection{Tuning the method} \label{tuning}
We performed $M=100$ replicates for each model. When $k$ is given, we compare
the results with those obtained using $k$-means, $k$-means(10) and MCLUST. Otherwise, if $k$ is unknown  we compared the results  with DBSCAN and MCLUST.  In order to apply CUBT we must fix the values for the parameters involved at each stage in the algorithm: for the maximal tree we used $minsize=5, 10$ or $15$ and $mindev=0.7$ or $0.9$; for the pruning stage $mindist=0.3$ or $0.5$ and  $\delta=0.2, 0.4$ or $0.6$ for the joining stage. For the cases where the number of clusters is stated, it is important to note that even though in almost every case the number of terminal nodes of the maximal tree is bigger than the number of groups, there is no warranty that this will happen. Then, if we are in that case we reduce $mindev$ in decreasing order, $0.6, 0.5,\dots$ until the maximal tree has at least $k$ terminal nodes.
For the cases where the number of clusters is not stated we consider the same parameters as in the previous case. To determine the number of clusters we must choose a threshold $\eta$. We consider the distances $d^\delta_{\widetilde{i},\widetilde{j}}$ defined in Section \ref{joining}
 that correspond to the tree that is the output of the pruning step, and fix $\eta$ as a low quantile of $d^\delta_{\widetilde{i},\widetilde{j}}$. Heuristically, low quantiles of $d^\delta_{\widetilde{i},\widetilde{j}}$ correspond to terminal nodes whose observations belong to the same clusters. The quantiles of $d^\delta_{\widetilde{i},\widetilde{j}}$ that determine $\eta$ chosen for M1 to M4 were $0.2,0.08,0.25$ and $0.15$ respectively.

%


Because we use synthetic data sets, we know the actual label of each observation, and it is thus reasonable to measure the goodness of a partition by computing ``the number of misclassified observations", which is analogous to the misclassification error for supervised classification procedures. We denote the original clusters $r=1,\ldots,R$. Let $y_1,...y_n$ be the group label of each observation, and $\hat{y}_1,...\hat{y}_n$ the class label assigned by the clustering algorithm. Let $\Sigma$ be the set of permutations over $\{1,...,R\}$. The misclassification error may then be expressed as:


\begin{equation}
MCE=\min_{\sigma \in\Sigma} \; \frac{1}{n} \sum_{i=1}^n \mathbf{1}_{\{y_i \ne \sigma(\hat{y}_i)\}}.
   \label{error}
\end{equation}


If the number of clusters is large, the assignment problem may be computed in polynomial time using Bipartite Matching and the Hungarian Method, \cite{Papad}. We use this algorithm only for M2 that has ten groups.

\subsection{Results}

First, we analyze the results of the simulation when the number of clusters is known.
Table \ref{resultsimul} shows the results obtained in our simulations. Except for Model 3, we varied the values of $\sigma$ to reflect the degree of overlapping between classes.
In order to study the robustness of the method with respects to the parameters selection we considered 36 different parameters configurations for our procedure. Since the results were practically the same, we present for each model the results for the best and the worst clustering allocation.
 We report the misclassification error obtained for each clustering procedure, namely CUBT for the best performance (CUBT (B)) and for the worst performance (CUBT (W)), $k$-Means, $k$-Means(10) and MCLUST (if $k$ is given). As expected for the first two models $k$-means(10) and MCLUST have a good performance, but both of them fail for M3. For the last model MCLUST performs poorly in both cases, $k$-means has an acceptable performance in both cases and $k$-means(10) always achieves a perfect cluster allocation. The results obtained by CUBT using different values for the parameters are practically the same, and in almost all cases the results for CUBT lie between those of $k$-Means and $k$-Means(10), except for Model 3 where it has a better performance and for M4 when $\sigma=0.05$ (where the misclassification rate of CUBT(W) is larger than the misclassification rate of $k$-means). If we compare the results with those of MCLUST we may say that for M1 they have a similar performance and that for M2, M3 and M4 the performance of CUBT is better.

\begin{table}[h]
\begin{center}
\begin{tabular}{|l|lllll|}
\hline
Sigma $\left( \sigma \right) $ & CUBT (W) & CUBT (B) & k-Means &
k-Means(10) & MCLUST \\ \hline

\multicolumn{6}{|c|}{Model 1} \\ \hline
$0.11$ & $0$ & $0$ & $0.12$ & $0$  & $0$ \\
$0.13$ & $7e-03$ & $0$ & $0.12$ & $0$  & $0$\\
$0.15$ & $1e-03$ & $1e-04$ & $0.11$ & $0$  & $0$\\
$0.17$ & $1e-03$ & $2e-04$ & $0.07$ & $5e-05$  & $3e-05$\\
$0.19$ & $2e-03$ & $3e-04$ & $0.06$ & $3e-04$  & $2e-04$\\\hline
\multicolumn{6}{|c|}{Model 2} \\\hline
$0.7$ & $0$ & $0$ & $0.11$ & $0.01$  & $0.04$ \\
$0.75$ & $0$ & $0$ & $0.10$ & $0.01$  & $0.05$\\
$0.8$ & $4e-04$ & $2e-04$ & $0.10$ & $0.01$  & $0.06$ \\
$0.85$ & $0.004$ & $0.002$ & $0.08$ & $0.01$  & $0.07$\\
$0.9$ & $0.05$ & $0.04$ & $0.07$ & $0.01$  & $0.08$\\ \hline
\multicolumn{6}{|c|}{Model 3} \\ \hline
$$ & $0$ & $3e-04$ & $0.47$ & $0.47$  & $0.25$ \\\hline
\multicolumn{6}{|c|}{Model 4} \\\hline
$0.03$ & $0$ & $0$ & $0.11$ & $0$  & $0.65$ \\
$0.05$ & $0.16$ & $0.05$ & $0.12$ & $0$  & $0.65$ \\

\hline\hline
\end{tabular}%
\caption{Simulation results for models M1 to M4.\label{resultsimul}}
\end{center}
\end{table}

Now we proceed to analyze the results of the simulation when $k$ is unknown. In Table \ref{gruposbien} we report the number of times that the procedure chooses the correct number of groups. The number of misclassified observations for CUBT and MCLUST, when the number of clusters is chosen properly, is the same as in the previous analysis. CUBT has a very good performance for M1, M3 and the first case of M4, and for the others situations the best performance is very good but it strongly depends on the choices of the other parameters. However, is important to note that in those cases identifying correctly  the clusters is a very difficult problem and the other methods were not able to do it. For M2 if $\sigma=0.7$ DBSCAN in nine opportunities identifies ten groups (in those cases all the observations are classified correctly) and in the rest of the replicates it identifies fewer groups, if the group overlapping is bigger it is not able to separate the groups, for $\sigma=0.75$ it always finds less than five groups and on the rest of the cases it always finds only one group. For M1 DBSCAN  has a very good performance identifying the number of groups and in those cases there are not misclassified observations except for the case of $\sigma=0.19$ that the misclassification rate is 0.027. For M3 and M4 whenever it identifies correctly the number of groups it also allocates the observations in the right group. MCLUST has an outstanding performance for M1, but it fails in all the other models. For M2 it always identifies one group and for M3 97 times finds nine groups and the rest of the times eight clusters. Finally for M4 even though it always finds three groups it fails in the allocation of the observations, which is consistent with the results found when the number of clusters was previously stated.
\begin{table}[h]
\begin{center}
\begin{tabular}{|l|llll|}
\hline
Sigma $\left( \sigma \right) $ & CUBT (W) & CUBT (B) & DBSCAN & MCLUST \\ \hline

\multicolumn{5}{|c|}{Model 1} \\ \hline
$0.11$ & $81$ & $85$ & $92$ & $99$  \\
$0.13$ & $75$ & $84$ & $81$ & $100$ \\
$0.15$ & $76$ & $87$ & $74$ & $100$ \\
$0.17$ & $79$ & $92$ & $43$ & $100$ \\
$0.19$ & $85$ & $88$ & $38$ & $100$ \\\hline
\multicolumn{5}{|c|}{Model 2} \\\hline
$0.7$ & $24$ & $90$ & $9$ & $0$  \\
$0.75$ & $34$ & $90$ & $0$ & $0$ \\
$0.8$ & $42$ & $90$ & $0$ & $0$  \\
$0.85$ & $58$ & $83$ & $0$ & $0$ \\
$0.9$ & $51$ & $78$ & $0$ & $0$ \\ \hline
\multicolumn{5}{|c|}{Model 3} \\ \hline
$$ & $94$ & $98$ & $76$ & $0$  \\\hline
\multicolumn{5}{|c|}{Model 4} \\\hline
$0.03$ & $100$ & $100$ & $100$ & $100$ \\
$0.05$ & $2$ & $98$ & $26$ & $100$ \\
\hline\hline
\end{tabular}%
\caption{Number of times that the different procedures choose the correct number of groups for  M1 to M4.\label{gruposbien}}
\end{center}
\end{table}




\subsection{A comparison between CART and CUBT}

In the following simple example, we compare the tree structure obtained by CART using the complete training sample (observations plus the group labels) with that obtained by CUBT considering only the training sample without the labels. We generated three sub--populations in a two-dimensional variable space. The underlying distribution of the vector $\mathbf{\mathbf{X}=}( X_{1},X_{2})$
is a bivariate normal distribution in which the variables  $X_{1}$ and $X_{2}$ are independent, with their distributions for the three groups being given by $X_{1} \sim \mathcal{N}\left( 0,0.03\right) ,  \ \mathcal{N}\left( 2,0.03\right) , \ \mathcal{N}\left( 1,0.25\right)$,
$X_{2}\sim \mathcal{N}\left( 0,0.25\right), \ \mathcal{N}\left( 1,0.25\right) , \ \mathcal{N}\left(2.5,0.03\right)$.

\begin{figure}[!ht]
  \includegraphics[width=10cm]{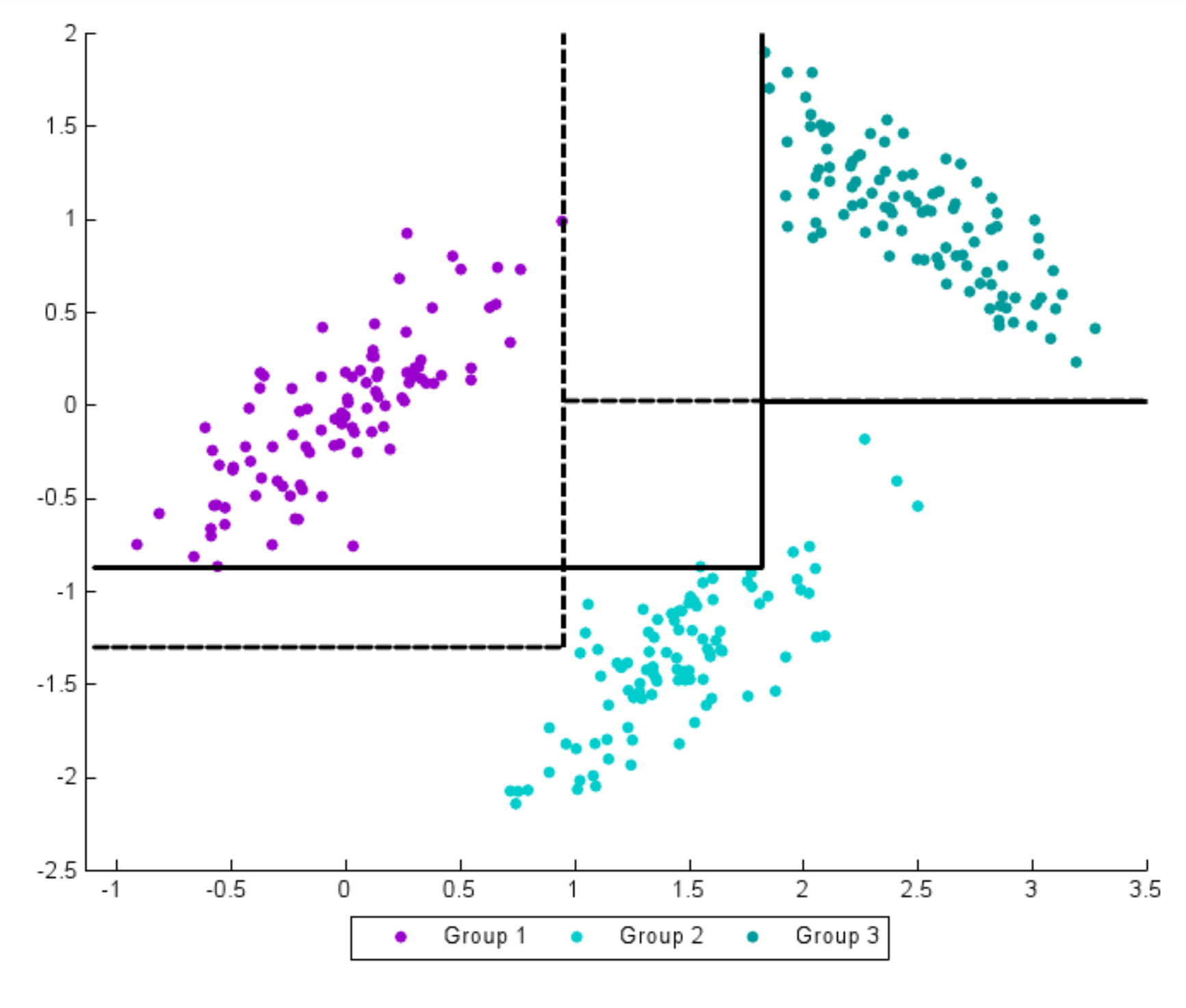}
\vspace{-15pt}
\caption{Plot of the partitions of the space for a generated data set. The solid lines indicate the partition for CUBT and the dashed lines the partition for CART}
\label{simulacion}
\end{figure}

The data were then rotated through $\pi /4$. One of the difficulties was that the optimal partitions are not parallel to the axis. Figure \ref{simulacion} shows the partitions obtained using CART and CUBT, for a single sample of size 300.
We performed 100 replicates, and in each case generated a training sample of size 300, whereby every group was of the same size. We then computed the ``mean misclassification rate" with respect to the true labels. For CUBT, the value was 0.09, while for CART there were no classification errors because we used the same sample both for growing the tree and for classifying the observations.

\begin{figure}[!ht]
\begin{center}
\includegraphics[height=3.5cm,width=6cm]{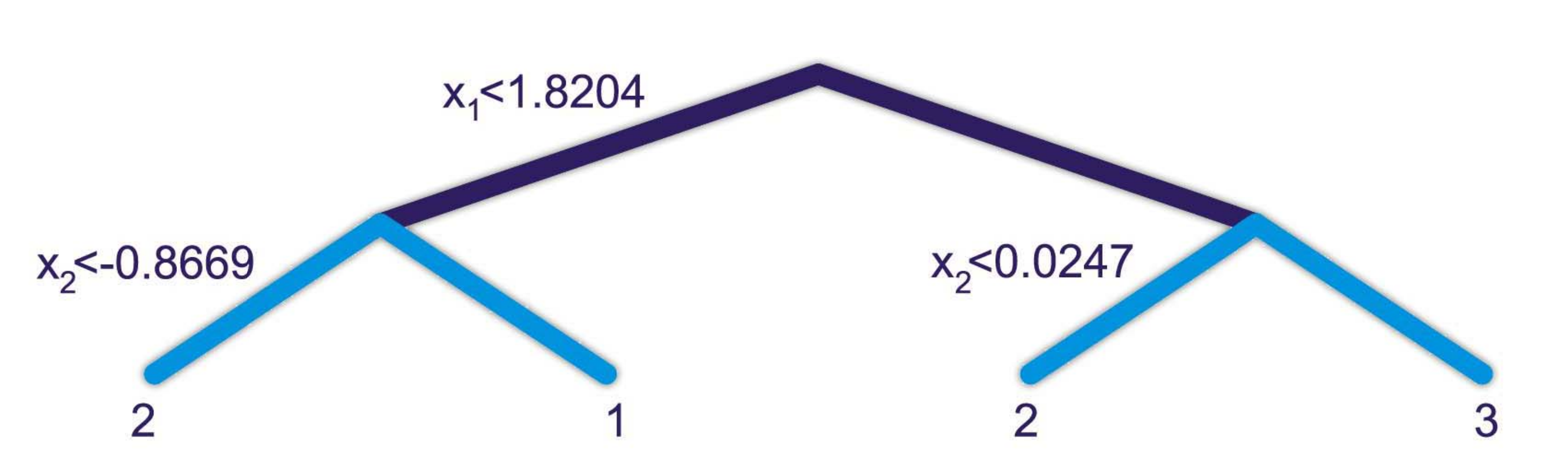} \hspace{1.5cm}
\includegraphics[height=3.5cm,width=6cm]{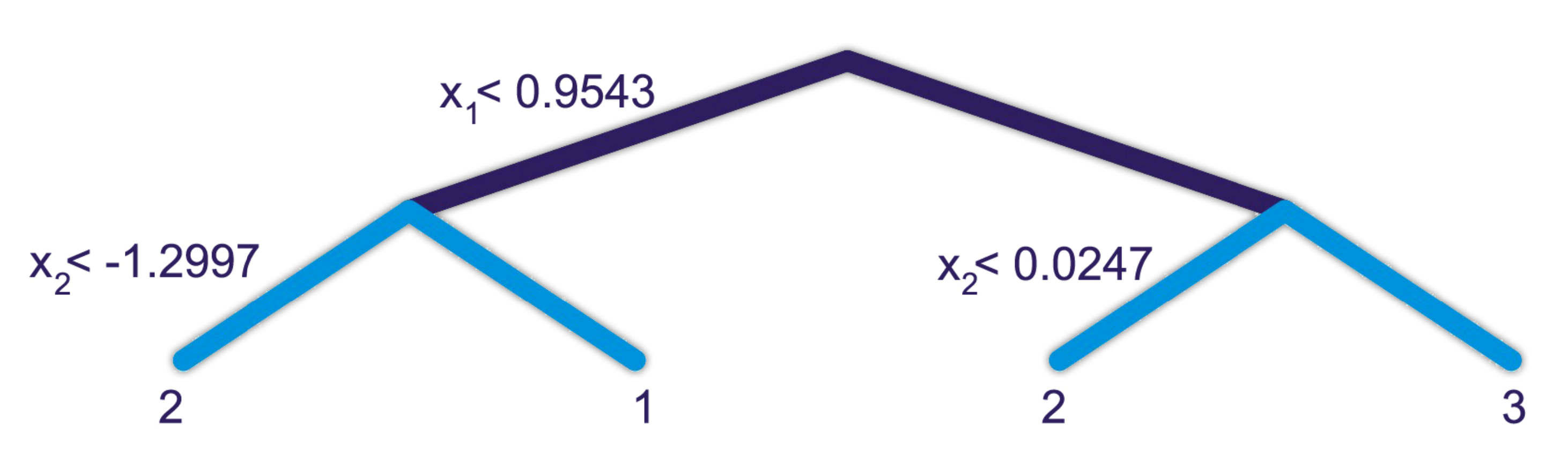}
\end{center}
\caption{ {\em Left:} Tree corresponding to CUBT . {\em Right:} Tree corresponding to CART. In both cases the left-hand branches indicate the smaller value of the partitioning variable.}  \label{treesimul}
\end{figure}

In order to compare our procedure with the traditional CART classification method, we obtained the binary trees for CART and CUBT. Both trees are shown in Figure ~\ref{treesimul}. It is noteworthy that the two structures are exactly the same, and that the different cutoff values in the different branches may be understood with the aid of Figure ~\ref{simulacion}, which corresponds to the same data set.

\section{A real data example. European Jobs}

In the following example, the data set describes the percentage of workers employed in different sectors of economic activity for a number of European countries in the year 1979 (this data set can be obtained on the website http://lib.stat.cmu.edu/DASL/Datafiles

\begin{figure}[!ht]
\begin{center}
\includegraphics[width=12cm]{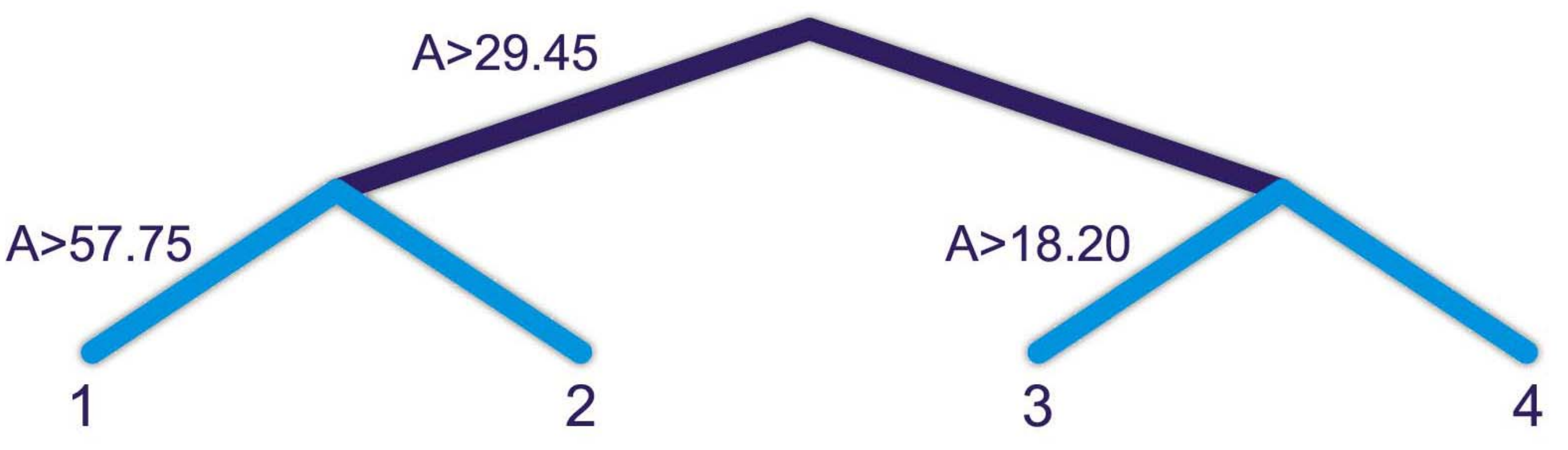}
\includegraphics[width=12cm]{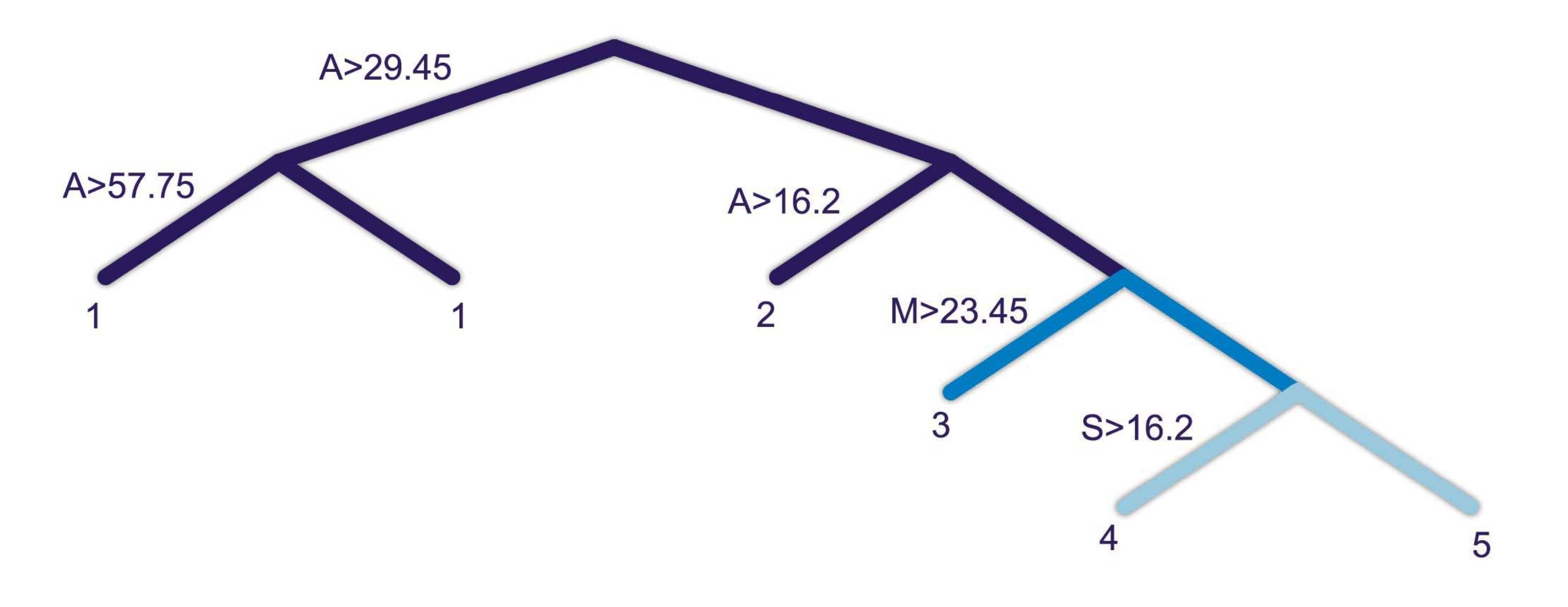}
\end{center}
\vspace{-15pt} \caption{Tree structure using four groups $\emph{top}$  and five groups $\emph{bottom}$, the left-hand branch shows the smaller values of the variable that is making the partition. }\label{agrick4}
\end{figure}

/EuropeanJobs.html). The categories are: agriculture (A), mining (M), manufacturing (MA), power supplies industries (P), construction (C), service industries (SI), finance (F), social and personal services (S) and transportation and communication (T). It is important to note that these data were collected during the cold war. The aim was to allocate the observations to different clusters, but the number of clusters is unknown. We must therefore study the data structure for a range of different numbers of clusters.
We first consider a four-group structure. In this case, a single variable (the percentage employed in agriculture) determines the tree structure. The four groups are shown in Table ~\ref{cabt4g}, and the corresponding tree is plotted on the top panel of Figure \ref{agrick4}. In the tree, the highest value of A corresponds to Turkey, which is an outlier and conforms to a single cluster of observations, which may be explained in terms of its social and territorial proximity to Africa.
The countries that make up groups 2 and 3 are those that were either under communist rule or those that were experiencing varying degrees of political upheaval; Spain, for example was making adjustments after the end of Franco's regime. The countries of Group 2 were poorer than those of Group 3. Finally, Group 4 had the lowest percentage of employment in agriculture, and the countries in this group were the most developed and were not under communist rule, with the exception of East Germany.
Using $k$-means we get the following clusters. Turkey and Ireland are isolated in one group each, Greece, Portugal, Poland, Romania and Yugoslavia form another group, and the remaining countries form a fourth cluster.

\begin{table}[!ht]
\begin{tabular}{llll}
\hline
Group 1 & Group 2 & Group 3 & Group 4 \\ \hline
Turkey & Greece & Ireland & Belgium \\
& Poland & Portugal & Denmark \\
& Romania & Spain & France \\
& Yugoslavia & Bulgaria & W. Germany \\
&  & Czechoslovakia & E. Germany \\
&  & Hungary & Italy \\
&  & USSR & Luxembourg \\
&  &  & Netherlands \\
&  &  & United Kingdom \\
&  &  & Austria \\
&  &  & Finland \\
&  &  & Norway \\
&  &  & Sweden \\
&  &  & Switzerland \\ \hline
\end{tabular}%
\caption{CUBT clustering structure using four groups. \label{cabt4g}}
\end{table}

If we use five clusters instead of four, the main difference is that Group 4 of the original partition is divided into two subgroups (4 and 5), and the variables that explain these partitions are the percentages employed in mining and agriculture. The other groups of the original partition remain stable.

If a five-cluster structure is considered via the $k$-means algorithm, Turkey and Ireland are then each isolated in single groups, and Greece, Portugal, Poland, Romania and Yugoslavia form another group, as in the four-cluster structure. Switzerland and East and West Germany make up a new cluster.

\begin{table}[!ht]
\begin{tabular}{lllll}
\hline
Group 1 & Group 2 & Group 3 & Group 4 & Group 5\\ \hline
Turkey & Greece & Ireland & Belgium & W. Germany\\
& Poland & Portugal & Denmark & E. Germany \\
& Romania & Spain & France & Switzerland\\
& Yugoslavia & Bulgaria & Italy & \\
&  & Czechoslovakia & Luxembourg & \\
&  & Hungary & Netherlands & \\
&  & USSR & United Kingdom \\
&  &  & Austria & \\
&  &  & Finland & \\
&  &  & Norway &\\
&  &  & Sweden &\\ \hline
\end{tabular}%
\caption{\label{cabt5g}CUBT clustering with five groups.}
\end{table}


\section{Concluding Remarks}

We have herein presented a new clustering method called CUBT which shares some ideas with the well known classification and regression trees, defining clusters in terms of binary rules over the original variables. Like CART, our method may be very attractive and useful in a number of practical applications.
 Because the tree structure makes use of the original variables, it helps to determine those variables that are important in the conformation of the clusters. Moreover, the tree allows the classification of new observations. In our approach, a binary tree is obtained in three stages. In the first stage, the sample is split into two sub-samples, thereby reducing the heterogeneity of the data within the new sub-samples according to the objective function $R(\cdot)$. The procedure is then applied recursively to each sub-sample. In the second and third stages, the maximal tree obtained at the first stage is pruned using a dissimilarity criterion, first applied to the adjacent nodes and then to all the terminal nodes.   The algorithm is simple and requires a reasonable computation time. There are no restrictions on the dimension of the data used. Our method is consistent under a set of quite general assumptions, and produces quite good results with the simulated examples that we considered here, as well as for an example that used real data.

The algorithm depends on several parameters, and an optimal way to choose them is beyond the scope of this paper.  We herein propose some advice in order choose them in practice. The splitting stage depends on two parameters, \textit{mindev} and \textit{minsize}. The later, $\textit{mindev}\in (0,1)$ represents the percentage of the deviance of the whole data set ($R(\mathcal{S})$) the algorithm requires to split a group (if the reduction of its deviance is less than $\textit{mindev}\times R(\mathcal{S})$ the group is split in two subgroups). Our experience indicates that values between 0.7 and 0.9 give a sufficiently large partition. The former indicates the minimum cluster size that the user admits, if he has some information beforehand it could be provided and taken into account by the algorithm otherwise the default value should be 1.  The pruning step includes also two parameters $\delta$ and $\epsilon$. In population terms, it suffices that $\epsilon$ be smaller than the distance among the supports of two disjoint clusters. If the user could not provide an input for this parameter the default could be zero, which correspond to skip the pruning step and go directly to the joining step, in that case one would probably obtain a larger tree, but the final clustering allocation would not change and the results given in Theorem \ref{theo2} still hold. The parameter $\delta$ is just a way to deal with some possible presence of outliers and as a default $\delta=0.2$ can be used. Finally, if the number of clusters is given the final stage does not require any parameter, otherwise one parameter, $\eta$, should be provided. The way we suggest to choose this parameter is given in Section \ref{tuning}.

A more robust version of our method could be developed by substituting the objective function $cov(X_T)$ in equation (\ref{deviance}) for a more robust covariance functional $ robcov(X_T)$ (see e.g. Maronna et al.\cite{Maronna} Chapter 6 for a review), and then proceeding in the same manner as described herein. However, a detailed investigation of this alternative approach lies beyond the scope of the present study.

Even though we focussed our paper on the treatment of interval data, the procedure can be extended to other settings as long as a covariance structure and a dissimilarity measure can be defined. For example, one could use Gini's definition of variance for categorical data, (\cite{gini}) and the Mean Character difference as a measure of dissimilarity. Other measures of dissimilarity for categorical data can be found in  Gan et al.\cite{gan}. However, a deeper study of these settings are beyond the scope of this paper.

\

\section{Acknowledgments}
Ricardo Fraiman and Marcela Svarc have been partially supported  by Grant  pict 2008-0921  from ANPCyT (Agencia Nacional de Promoción Cient\'{i}fica y Tecnol\'{o}gica), Argentina.

 \section{Appendix}

  \subsection{Proof of Lemma \ref{lem1} }

  We first observe that because  $t_l$ and $t_r$ are disjoint, $\mathbb{E}(X_{t_l \cup t_r}) =  \gamma  \mu_l + (1 -  \gamma)  \mu_r$, where $\gamma = P(\mathbf{X} \in t_l \vert \mathbf{X} \in t_l \cup t_r)$.
   Given that $j = 1, \ldots, p$,
   we use $ M^{(j)}_{2i} =  \int_{t_i} x(j)^2 dF(x),  \   \
   i=l,r$, where $F$ stands for the distribution function of the
   vector $\mathbf{X}$.

   It then follows that $\mathbb{E}(X_{t_l \cup t_r}(j)^2) = \gamma M^{(j)}_{2l}+ (1 -  \gamma)M^{(j)}_{2r}$ , and therefore that
    $$
    var( X_{t_l \cup t_r}(j)) = \gamma var(X_{t_r}(j)) + (1-\gamma) var(X_{t_r}(j)) + \gamma (1-\gamma )( \mu_l(j) - \mu_r(j))^2.
    $$
    By summing the terms in $j$ we get the desired result.

\subsection{Proof of Theorem \ref{theo1} }

Let $\mathbb{T}$ be the family of polygons in $ \mathbb{R}^p$ with faces orthogonal to the axes, and fix $i \in \{1, \ldots, p\}$ and $t  \in \mathbb{T}$. For $a \in \mathbb{R}$ denote by $t_l =   \{ x \in t: x(i) \leq a\}$ and $t_r = t \setminus t_l$. We define $r(t,i,a) = R(t) - R(t_l) - R(t_r)$ and $r_n(t,i,a) = R_n(t) - R_n(t_l) - R_n(t_r)$, the corresponding empirical version.

We start showing the uniform convergence
\begin{equation}
\label{convunif}
\sup_{a \in \mathbb{R}} \sup_{t \in \mathbb{T}}\vert r_n(t,i,a)-r(t,i,a)\vert \to 0 \ \ a.s.
\end{equation}

By Lemma \ref{lem1},

\begin{equation}
\label{nuevoobjetivo}
\alpha_t r(t,i,a) = \alpha_{t_l}\alpha_{t_r} \Vert \mu_l(a) - \mu_r(a)\Vert^2,
\end{equation}
where $\alpha_A = P(\mathbf{X} \in A)$ and  $ \mu_j(a) = \mathbb{E}(X_{t_j}), j=l,r$. Then, the pairs $(i_{jn}, a_{jn})$ and $(i_{j}, a_{j})$ are the arguments that maximise the right-hand side of equation  (\ref{nuevoobjetivo}) with respect to the measures $P_n$ and $P$ respectively.  We observe that the right-hand side of equation (\ref{nuevoobjetivo}) equals

\begin{equation}
\alpha_{t_r}  \int_{t_l}  \Vert x \Vert ^2dP(x) + \alpha_{t_l}  \int_{t_r}  \Vert x \Vert ^2dP(x) - 2 \langle\int_{t_l} xdP(x) , \int_{t_r} xdP(x)\rangle.
\end{equation}

It order to prove equation (\ref{convunif}) it is sufficient to show that:
\begin{enumerate}
\item  $ \sup_{a \in \mathbb{R}} \sup_{t \in \mathbb{T}}\vert P_n(t_j) - P(t_j)\vert \to 0 \ \ a.s. \ \ j=l,r$
\item  $ \sup_{a \in \mathbb{R}}\sup_{t \in \mathbb{T}}\vert \int_{t_j} \Vert x \Vert^2 dP_n(x) -  \int_{t_j} \Vert x \Vert^2 dP(x) \vert \to 0 \ \ a.s. \ \ j=l,r$
\item  $ sup_{a \in \mathbb{R}}\sup_{t \in \mathbb{T}} \vert \int_{t_j}\ x (i) dP_n(x) -  \int_{t_j}  x(i) dP(x) \vert \to 0 \ \ a.s. \ \ j=l,r, i=1, \ldots, p.$
\end{enumerate}
Since $\mathbb{T}$ is a Vapnik--Chervonenkis class, we have that (i) holds. Now observe that the conditions for uniform convergence over families of sets still hold if we are dealing with signed finite measures. Therefore if we consider the finite measure $ \Vert x \Vert ^2dP(x)$ and the finite signed measure given by $x(i)dP(x)$ we also have that (ii) and (iii) both hold.

\
Since $$\lim_{a \to \infty}  \alpha_{t_l}\alpha_{t_r} \Vert \mu_l(a) - \mu_r(a)\Vert ^2= \lim_{a \to -\infty}  \alpha_{t_l}\alpha_{t_r} \Vert \mu_l(a) - \mu_r(a)\Vert ^2= 0, $$
we have that
$\inf{\{argmax_{a \in \mathbb{R}} r_n(t,i,a)\}} \to  \inf{\{argmax_{a \in \mathbb{R}} r(t,i,a)\}}$ a.s.

In the first step of the algorithm, $t=\mathbb{R}^p$ and we obtain $ i_{n1} = i_1$ for $n$ large enough and $a_{n1} \to a_1$ a.s. In the next step, we observe that the empirical procedure begins to work with $t_{nl}$ and $t_{nr}$, while the population algorithm will do so with $t_{l}$ and $t_{r}$. However, we have that

$$
 \sup_{a \in \mathbb{R}} \vert  r_n(t_{nj}, i, a) - r(t_j, i, a) \vert \leq
 $$
\begin{equation}
\label{final}
 \sup_{a \in \mathbb{R}} \sup_{t \in \mathbb{T}}\vert  r_n(t_{nj}, i, a) - r(t_{nj}, i, a) \vert +
 \sup_{a \in \mathbb{R}} \vert  r(t_{nj}, i, a) - r(t_j, i, a) \vert ,
 \end{equation}

for $j=l,r$.

We already know that the first term on the right hand side of equation(\ref{final}) converges to zero almost surely. In order to show that the second term also converges to zero, it is sufficient to show that
\begin{enumerate}
\item  $ \sup_{a \in \mathbb{R}} \vert P(t_{nj})- P(t_j)\vert \to 0 \ \ a.s. \ \ j=l,r$
\item  $ \sup_{a \in \mathbb{R}}\vert \int_{t_j} \Vert x \Vert^2 dP(x) -  \int_{t_{nj} } \Vert x \Vert^2 dP(x) \vert \to 0 \ \ a.s. \ \ j=l,r$
\item $ sup_{a \in \mathbb{R}} \vert \int_{t_j}\ x (i) dP(x) -  \int_{t_{nj}}  x(i) dP(x) \vert \to 0 \ \ a.s. \
 \ j=l,r, \  i=1, \ldots, p,$
 \end{enumerate}
which follows from the assumption that $\Vert x \Vert ^2 f(x)$ is bounded. This concludes the proof since $minsize/n \to \tau$.

\

\subsection{Proof of Theorem \ref{theo2} }

We need to show that we have consistency in both steps of the backward algorithm.

(i) Convergence of the pruning step. Let  $\{t^*_{1n}, \ldots, t^*_{mn}\}$ be the output of the forward algorithm. The pruning step partition of the algorithm converges to the corresponding population version from
\begin{itemize}
\item the conclusions of Theorem \ref{theo1}.
\item the fact that the random variables $W_{lr} \ \ \tilde{d}_l, \tilde{d}_r$ are positive.
\item the uniform convergence of the empirical quantile function to its population version.
\item the Lebesgue dominated convergence Theorem.
\end{itemize}

The proof of (ii) is mainly the same as that for (i).







\end{document}